\documentclass[%
reprint,
superscriptaddress,
showpacs,preprintnumbers,
amsmath,amssymb,
aps,
prb,
floatfix,
]{revtex4-1}

\usepackage{graphicx}
\usepackage{dcolumn}
\usepackage{bm}
\usepackage{bbm}

\usepackage[
text={7.2in,8.7in},centering
]{geometry}
\usepackage[utf8]{inputenc}
\usepackage{bigints}
\usepackage{cleveref}
\crefname{equation}{Eq.}{Eqs.}
\usepackage[version=3]{mhchem} 
\usepackage{color}

\definecolor{pinegreen}{rgb}{0.33,0.5,0.2}

\begin{document} 

\title{Excitation Energies from Thermally-Assisted-Occupation Density Functional Theory: Theory and Computational Implementation} 

\author{Shu-Hao Yeh} 
\thanks{These authors contributed equally.} 
\affiliation{Institute of Chemistry, Academia Sinica, Taipei 11529, Taiwan} 
\affiliation{Department of Chemistry, National Taiwan University, Taipei 10617, Taiwan} 

\author{Aaditya Manjanath} 
\thanks{These authors contributed equally.} 
\affiliation{Institute of Chemistry, Academia Sinica, Taipei 11529, Taiwan} 

\author{Yuan-Chung Cheng} 
\affiliation{Department of Chemistry, National Taiwan University, Taipei 10617, Taiwan} 

\author{Jeng-Da Chai} 
\email[Corresponding author: ]{jdchai@phys.ntu.edu.tw} 
\affiliation{Department of Physics, National Taiwan University, Taipei 10617, Taiwan} 
\affiliation{Center for Theoretical Physics and Center for Quantum Science and Engineering, National Taiwan University, Taipei 10617, Taiwan} 

\author{Chao-Ping Hsu} 
\email[Corresponding author: ]{cherri@sinica.edu.tw} 
\affiliation{Institute of Chemistry, Academia Sinica, Taipei 11529, Taiwan} 

\date{August 1, 2020} 

\begin{abstract} 

The time-dependent density functional theory (TDDFT) has been broadly used to investigate the excited-state properties of various molecular systems. However, the current TDDFT heavily relies on outcomes from the corresponding ground-state density functional theory (DFT) calculations which may be prone to errors due to the lack of proper treatment in the non-dynamical correlation effects. Recently, thermally-assisted-occupation density functional theory (TAO-DFT) [J.-D. Chai, \textit{J. Chem. Phys.} \textbf{136}, 154104 (2012)], 
a DFT with fractional orbital occupations, was proposed, explicitly incorporating the non-dynamical correlation effects in the ground-state calculations with low computational complexity. 
In this work, we develop time-dependent (TD) TAO-DFT, which is a time-dependent, linear-response theory for excited states within the framework of TAO-DFT.   
With tests on the excited states of H$_{2}$, the first triplet excited state ($1^3\Sigma_u^+$) was described well, with non-imaginary excitation energies. TDTAO-DFT also yields zero singlet-triplet gap in the dissociation limit, for the ground singlet ($1^1\Sigma_g^+$) and the first triplet state ($1^3\Sigma_u^+$). In addition, as compared to traditional TDDFT, the overall excited-state potential energy surfaces obtained from TDTAO-DFT are generally improved and better agree with results from the equation-of-motion coupled-cluster singles and doubles (EOM-CCSD).

\end{abstract} 

\maketitle


\section{Introduction}
Over the past decades, Kohn-Sham density functional theory (KS-DFT)~\cite{DFT, KSDFT} has been extensively used in the prediction of various ground-state properties of solids as well as finite-sized molecules.~\cite{Jones2015,Becke2014,Mardirossian2017}
Its time-dependent 
(TD) extension, known as TDDFT~\cite{ALDA_fail_1, ALDA_success_2,Casida_chapter}, has been a popular approach for computing excited-state properties, including the absorption and emission spectra~\cite{TDTAO_preamble}, photochemical reactions~\cite{PES_TDDFT}, dynamics~\cite{MD_TDDFT}, energy and electron transfer~\cite{electron_transfer_TDDFT}, etc., due to its 
low computational cost and the availability of a plethora of computer codes in this area. The one-to-one correspondence between the TD density and the TD external potential was rigorously demonstrated by Runge and Gross in 1984 in their theorem~\cite{ALDA_fail_1}. The linear-response framework was further introduced ~\cite{ALDA_success_2,Casida_chapter}, which brought forth a paradigm shift in the simulation of excitations of quantum systems from a density-functional perspective\cite{Hirata1999,Stratmann1998,Hsu2001} and is the main reason behind the popularity of this method.

However, conventional TDDFT 
is derived from ground-state (GS) KS-DFT which is a single-determinant--based method. 
As a result, it can fail to describe the excited-state phenomena heavily governed by non-dynamical (or static) correlation, such as photochemistry processes involving photoinduced bond breaking, and problems associated with conical intersection~\cite{ALDA_success_2, TDTAO_preamble, Levine2006, Filatov2013}. A prototypical example is the bond dissociation process of the H$_{2}$ molecule. It is known that the excitation energy of the lowest triplet state of \ce{H2}, computed using conventional TDDFT~\cite{TDTAO_preamble}, would become imaginary beyond a H-H bond distance of 1.75~\AA, a phenomenon arising from a spin symmetry-breaking solution in the ground state~\cite{Baerends_2000, Casida_2000}, a typical characteristic of nondynamical correlation effects. In contrast, in wavefunction-based methods, the (nearly) degenerate determinants are considered on an equal footing when performing a self-consistent field (SCF) calculation, and this is the basis of multi-configuration (MC) SCF or complete active space (CAS) SCF-based methodologies. However, these methods can be prohibitively expensive for large systems, as their computational cost scales factorially with the size of active space. 

KS-DFT with proper exchange energy functionals may reasonably model systems with non-dynamical correlation, albeit at the expense of enormous computation efforts. For example, the works by Becke~\cite{becke_1, becke_2} and the works by Kong and coworkers~\cite{jing_kong_1, jing_kong_2, jing_kong_3} demonstrated parametric functionals which need to be solved self-consistently within the single-determinant framework. Although these works significantly improved the bond dissociation trends of simple diatomic molecules, compared to the Hartree-Fock theory, they still deviate appreciably at the bond dissociation limit compared to a full configuration interaction (FCI) calculation~\cite{jing_kong_1, jing_kong_2}. Moreover, the SCF associated with these functionals adds to the computational effort which can scale dramatically with the size of molecules. 

On the other hand, various approaches have been developed to cope with the non-dynamical correlation effects without the high computational cost of an exact exchange functional. The CAS-DFT model is one such method~\cite{CAS-DFT}, wherein some amount of correlation has been accounted for, by a density functional calculation. As a result, the dynamical correlation associated with the MC representation of the system might be ``double counted"~\cite{MCPDFT, MCPDFT_correction}. To mitigate this issue, the multi-configuration pair-density functional theory~\cite{MCPDFT, MCPDFT_correction} and multi-configuration range-separated DFT~\cite{TDMCPDFT,sharkas2012} were developed. While the former utilizes the so-called \textit{on-top pair-density functional}, the latter separates the electron interaction operator into short- and long-range parts which are treated with DFT and wavefunction theory, respectively.
Although the idea of using such a ``hybrid'' scheme seems to be an attractive prospect~\cite{MCPDFT, MCPDFT_correction, TDMCPDFT,sharkas2012}, they can be computationally demanding for increasing system sizes because of the initial generation of MC wavefunctions.

Another category of computational methods exists, which can cope with non-dynamical correlation with the additional advantage that they are low-cost methods. They include the spin-flip, ionization-potential, and electron-affinity based approaches which are aimed to start with a high-spin, with 1-less or 1-more electron single-determinant references such that the non-dynamical correlation problem is minimal~\cite{Stanton1994, Nooijen1998, Shao2003}. These approaches require a well-balanced treatment of the orbitals in the reference, and they can offer high-quality solutions in many cases. However, the requirement of balanced treatment of orbitals in the reference is not always feasible, and thus applications are limited.

In this regard, the thermally-assisted-occupation density functional theory (TAO-DFT)~\cite{TAODFT_1} was developed by Chai in 2012 to alleviate the formidable challenge of balancing the computational cost and simultaneously incorporating the non-dynamical correlation effects with reasonable accuracy. In contrast to traditional KS-DFT, the underlying principle of TAO-DFT is in the usage of fractional orbital occupations according to a given fictitious temperature ($\theta$), to effectively incorporate the different electronic configurations of a system. This approach ensures that some ``excitations'' in the form of fractional populations of electrons in the low-lying virtual orbitals are considered along with the GS of the system, similar to a multi-determinant expansion of the wavefunction. The inclusion of fractional occupancies is a computationally cheaper alternative to a multi-determinant expansion for accounting non-dynamical correlation effects. As a result, TAO-DFT has a computational cost similar to that of KS-DFT, which is $\mathcal{O}(N^{3\text{-}4})$. In TAO-DFT, the entropy contribution (e.g., see Eq.\ (26) of Ref.\citenum{TAODFT_1}), can reasonably capture the non-dynamical correlation energy of a system, which was discussed and numerically investigated in Ref.~\citenum{TAODFT_1}, even when the simplest local density approximation (LDA) XC energy 
functional is used. The XC energy functionals at the higher rungs of Jacob's ladder, such as the generalized-gradient approximation (GGA) \cite{TAODFT_2}, global hybrid \cite{TAODFT_3}, 
and range-separated hybrid \cite{TAODFT_3, TAODFT_5} XC energy functionals, can also be employed in TAO-DFT. Moreover, a self-consistent scheme that determines the fictitious temperature 
in TAO-DFT has been recently proposed to improve the performance of TAO-DFT for a wide range of applications \cite{TAODFT_4}. Since TAO-DFT is similar to KS-DFT in computational efficiency, 
TAO-DFT has been recently adopted for the study of the electronic properties of various nanosystems with pronounced radical 
nature \cite{TAODFT_5, TAODFT_6, TAODFT_7, TAODFT_8, TAODFT_9, TAODFT_10, TAODFT_11, TAODFT_12, TAODFT_13}. In particular, the electronic properties (e.g., singlet-triplet 
energy gaps, vertical ionization potentials, vertical electron affinities, fundamental gaps, and active orbital occupation numbers) of linear acenes and zigzag graphene nanoribbons (i.e., systems with polyradical character) obtained from TAO-DFT \cite{TAODFT_1, TAODFT_2, TAODFT_3, TAODFT_6} have been shown to be in reasonably good agreement with those obtained from other accurate electronic structure methods, such as the particle-particle random-phase approximation (pp-RPA) \cite{acene_weitao} XC energy functional in KS-DFT, the density matrix renormalization group (DMRG) algorithm \cite{acene_dmrg, zgnrs_dmrg}, the variational two-electron reduced density matrix (2-RDM) method \cite{acene_2rdm, acene_2rdm2}, and other high-level methods \cite{acene_others, acene_others2, acene_others3, acene_others4}.

\section{Ground-state reference: TAO-DFT}
In TAO-DFT~\cite{TAODFT_1}, the electron density is represented by the thermal equilibrium density of an auxiliary system of $N_\mathrm{e}$ non-interacting electrons at a fictitious temperature 
$\theta$ (in energy units): 
\begin{eqnarray}\label{TAO-dens} 
\rho(\mathbf{r}) = \sum_i f_i \, \phi^*_{i}(\mathbf{r}) \phi_{i}(\mathbf{r}). 
\end{eqnarray} 
Here, $f_i$ (a value between 0 and 1) is the fractional occupation number of the $i^\text{th}$ orbital $\phi_i$, and is given by the Fermi-Dirac distribution function 
\begin{eqnarray}
f_{i} = \Big\{1+\exp\big{[}(\varepsilon_{i}-\mu)/\theta\big{]}\Big\}^{-1}, 
\end{eqnarray} 
where $\mu$ is the chemical potential for electrons, and is determined by $\sum_i f_i = N_\mathrm{e}$ for a given $\theta$, orbital energies $\{\varepsilon_i\}$, and total electron number 
$N_\mathrm{e}$. This choice for the fractional occupation function and the corresponding one-particle density matrix has been extensively used in other methods such as finite-temperature DFT (FT-DFT)~\cite{Mermin1965} and floating occupation molecular orbital-complete active space configuration interaction (FOMO-CASCI)~\cite{Martinez2010}. With this {\it assisted} occupation number and generalized density expression, the total ground-state energy functional can be written as 
\begin{equation}\label{TAOenergy} 
E_\text{G}[\rho] = T_{\text{TAO}[\{f_{i},\phi_{i}\}] }+ V_{\text{ext}}[\rho] + E_{\text{Hxc}}^\text{KS} + E_{\theta}[\rho], 
\end{equation} 
where $T_{\textrm{TAO}}$ is the kinetic free energy functional of non-interacting electrons (equivalent to $A_s^\theta$ as defined in  Eq.~(24) of Ref.~\citenum{TAODFT_1}), $V_\text{ext}[\rho]$ is the energy functional of the external potential (or nuclei potential), $E_{\text{Hxc}}^{\text{KS}}$ is the sum of Hartree and XC energy functionals in KS-DFT, and $E_\theta$ is the $\theta$-dependent energy functional \cite{TAODFT_1}. Alternatively (to the original derivation \cite{TAODFT_1}), from \Cref{TAOenergy}, upon performing the functional derivatives with respect to the orbitals ($\phi_i$), we can also obtain the SCF equations in TAO-DFT: 
\begin{equation} 
\bigg[-\frac{1}{2}\nabla_\mathbf{r}^2 + v_\text{ext}(\mathbf{r}) + v_\text{Hxc}^\text{KS}(\mathbf{r}) + v_\theta(\mathbf{r}) \bigg]\phi_i(\mathbf{r}) = \varepsilon_i\phi_i(\mathbf{r}), 
\end{equation} 
where $v_\text{ext}$, $v_\text{Hxc}^\text{KS}$, and $v_\theta$ are the potentials (or functional derivatives) of corresponding energy functionals (i.e., $V_{\text{ext}}[\rho]$, $E_{\text{Hxc}}^\text{KS}$, and $E_{\theta}[\rho]$, respectively) in \Cref{TAOenergy}, and $\{\phi_i\}$ and $\{\varepsilon_i\}$ are the TAO orbitals and orbital energies, respectively, which can be solved self-consistently through SCF. The algorithm is similar to KS-DFT, with the only differences being the $v_\theta(\mathbf{r})$ term in the Hamiltonian and the determination of chemical potential $\mu$, making this approach attractive and easy in implementation. We have provided a variational perspective of TAO-DFT in Appendix~\ref{TAO-DFT_derivation}, which complements the derivation in Ref.~\citenum{TAODFT_1}.

\section{Excited state theory: TDTAO-DFT}
\subsection{Mathematical Formalism}
In the present work, we propose TDTAO-DFT, which is a time-dependent  linear-response  theory for TAO-DFT, allowing excitation energy calculation using Casida's formulation~\cite{Casida_chapter}, within the framework of TAO-DFT. 
In TDTAO-DFT, the TD density is given by 
\begin{eqnarray}
\rho(\mathbf{r},t) = \sum_p f_p \, \phi^*_{p}(\mathbf{r},t) \phi_{p}(\mathbf{r},t),
\label{taodensity}
\end{eqnarray} 
where $\phi_{p}(\mathbf{r},t)$ are the TD orbitals (for the fictitious particles), and $f_p$ are the corresponding fractional occupation numbers, which are assumed to be {\it time-independent}, and their values are taken from those obtained from the corresponding ground-state TAO-DFT calculation (Eq.~\ref{TAO-dens}). In order to facilitate the mapping between the original interacting system of electrons moving under the influence of a TD external potential and the auxiliary system of non-interacting particles, an action variational principle in TAO-DFT should be established. Following the variational principle, the TD effective potential for the non-interacting TAO system can be partitioned into the following parts:
\begin{equation} 
v_{\mathrm{eff}}^\mathrm{TAO}(\mathbf{r},t) = v_{\mathrm{ext}}(\mathbf{r},t) + v^{{\mathrm{TAO}}}_{\mathrm{Hxc}\theta}[\rho](\mathbf{r},t) 
\end{equation} 
where $v^{{\mathrm{TAO}}}_{\mathrm{Hxc\theta}}$ is the functional derivative of the $\mathrm{Hxc\theta}$-action, which contains the time-dependent Hartree potential, exchange-correlation potential, and the $\theta$ potentials for the fractional occupation. Further details are included in Appendix~\ref{TDTAO_variational_principle} accompanying this work.

Similar to conventional TDDFT,
with the equality connecting the effective potential and the functional derivative of TD action, an equation of motion for TDTAO-DFT can be expressed as 
\begin{eqnarray} 
i\frac{\partial}{\partial t}
\phi_p(\mathbf{r},t)
&=&
\bigg{[}-\frac{1}{2} \nabla_{\mathbf{r}}^2
+v_{\mathrm{ext}}(\mathbf{r},t)
+v^{{\mathrm{TAO}}}_{\mathrm{Hxc}\theta}[\rho](\mathbf{r},t)
\bigg{]} 
\phi_{p}(\mathbf{r},t)
\nonumber \\
&=& \hat{F}(t)  \,\phi_{p}(\mathbf{r},t).
\label{TDTAO_TDKS}
\end{eqnarray}
We note that $v^{{\mathrm{TAO}}}_{\mathrm{Hxc}\theta}[\rho](\mathbf{r},t)$ is also a TD generalization of the potential associated with the Hartree, exchange, correlation and 
$\theta$-functionals in GS TAO-DFT. The equation of motion is reformulated in terms of the one-particle density matrix $\mathbf{P}(t)$~\cite{TDTAO_preamble}: 
\begin{eqnarray}
i\frac{\partial}{\partial t} \mathbf{P}(t)=
\big{[} \mathbf{F}(t),\mathbf{P}(t)\big{]}
\label{matrix_TDTAO_TDKS}
\end{eqnarray}
where $ \mathbf{F}(t)$, the time-dependent ``Fock matrix'', is the matrix representation of the one-particle operator ($\hat{F}$) in Eq.~\ref{TDTAO_TDKS}. The general time-evolution of 
the state of a system is given by: 
\begin{eqnarray}
\mathbf{P}(t) &=& \mathbf{P}^{\circ} + \delta\mathbf{P}(t)\,\,\mathrm{and} \\
\mathbf{F}(t) &=& \mathbf{F}^{\circ} + \delta\mathbf{V}_{\textrm{ext}}(t) + \delta\mathbf{F}_{\mathrm{Hxc}\theta}[\mathbf{P}](t),
\end{eqnarray}
where $\mathbf{P}^{\circ}$ and $\mathbf{F}^{\circ}$ denote the initial conditions for solving Eq.~\ref{matrix_TDTAO_TDKS}, $\delta\mathbf{P}(t)$, $\delta\mathbf{V}_\mathrm{ext}(t)$, and $\delta\mathbf{F}^\text{e-e}[\mathbf{P}](t)$ are the time-dependent changes in the matrices of density, external field, and the electron-electron interaction, respectively, in the system. The initial state (at $t=t_0$) is commonly considered to be the unperturbed GS of the system for convenience. In terms of the GS TAO orbitals: 
\begin{eqnarray}
P^{\circ}_{pq}=\delta_{pq}\cdot f_p;~ 
F^{\circ}_{pq}=\delta_{pq}\cdot\varepsilon_p
\label{ground-state-dens-fock}
\end{eqnarray}

If the electronic eigenspectrum of a system is desired, the amplitude of the change in the external field $|\delta\mathbf{V}_\mathrm{ext}(t)|$ is assumed to be infinitesimally small~\cite{TDTAO_preamble, ALDA_fail_1, ALDA_success_2, Casida_chapter}. It is therefore suitable to consider a linear response relation between $\delta\mathbf{F}_{\mathrm{Hxc}\theta}[\mathbf{P}](t)$ and $\delta\mathbf{P}(t)$. Using the GS TAO orbital basis this can be obtained as 
\begin{eqnarray}
&&\delta F_{rs}^{\mathrm{Hxc}\theta}(t) = \sum_{pq}  \int  \mathrm{d} \tau
\Bigg{(}\frac{\, \delta F_{rs}^{\mathrm{Hxc}\theta}(t) \,}{\delta {P}_{pq}(\tau)}\Bigg{)}
\,\delta P_{pq}(\tau).
\label{linear_response_relation}
\end{eqnarray}
Employing the time-domain Fourier transformations 
\begin{eqnarray}
&&\delta P_{qr}(\omega)=\int  \mathrm{d} t \,\text{e}^{-i \omega t}  \big{[}\delta P_{qr}(t)\big{]} \\
&&\delta V_{qr} (\omega) = \int \mathrm{d} t \,\text{e}^{-i \omega t}  \big{[}\delta V_{qr}(t)\big{]} \\
&&\frac{\, \delta F_{rs}^{\mathrm{Hxc}\theta} \,}{\delta {P}_{pq}}(\omega)=\int  \mathrm{d} t \,\text{e}^{-i \omega (t-\tau)}  \bigg{[}\frac{\, \delta F_{rs}^{\mathrm{Hxc}\theta}(t) \,}{\delta {P}_{pq}(\tau)}\bigg{]} 
\end{eqnarray}
one could recast Eq.~\ref{matrix_TDTAO_TDKS}
into 
\begin{eqnarray}
\sum_q 
&\Bigg{[}&
F^{\circ}_{pq}\cdot\delta P_{qr}(\omega)
-
\delta P_{pq}(\omega)\cdot F^{\circ}_{qr} \nonumber \\
&&+
\bigg{(}
\delta V_{pq}(\omega)+
\sum_{st} \Bigg{(}\frac{\delta F_{pq}^{\mathrm{Hxc}\theta}}{\delta {P}_{st}}(\omega)\Bigg{)}\cdot
\delta P_{st}(\omega)
\bigg{)}
P^{\circ}_{qr}
\nonumber  \\
&&-\,
P^{\circ}_{pq}
\bigg{(}
\delta V_{qr}(\omega)+
\sum_{st} \Bigg{(}\frac{\delta F_{qr}^{\mathrm{Hxc}\theta}}{\delta {P}_{st}}(\omega)\Bigg{)}\cdot
\delta P_{st}(\omega)
\bigg{)}
\Bigg{]} \nonumber \\
&=&
\omega\cdot\delta P_{pr}(\omega)
\end{eqnarray}
by neglecting all second-order (or higher) terms. Upon invoking the GS definitions in Eq.~\ref{ground-state-dens-fock} and assuming all $\delta V_{pq}(\omega)$ to be infinitesimally small, the corresponding working equation becomes
\begin{eqnarray}
(\varepsilon_p-\varepsilon_r)\delta P_{pr}(\omega)&-&(f_p-f_r)
\Bigg{[}
\sum_{st} \bigg{(}\frac{\delta F^{\mathrm{Hxc}\theta}_{pr}}{\delta {P}_{st}}(\omega)\bigg{)}
\delta P_{st}(\omega)
\Bigg{]} \nonumber \\
&=& \omega\cdot\delta P_{pr}(\omega).
\end{eqnarray}
A conventional linear-response relation (which is the \textit{inverse} of Eq.~\ref{linear_response_relation})~\cite{ALDA_success_2,ullrich2012time} gives the TD density-density response function. The details of this derivation are provided in Appendix~\ref{density-density_response}.

Similar to conventional TDDFT, we apply the adiabatic approximation to the $\text{xc}\theta$-kernel (i.e., the $\text{xc}\theta$-kernel is assumed to be frequency-independent)~\cite{TDTAO_preamble, Casida_chapter, ALDA} 
\begin{eqnarray} 
\frac{\delta F^{\mathrm{Hxc}\theta}_{pr}}{\delta {P}_{st}}(\omega)&\approx &
\frac{\delta F^{\mathrm{Hxc}\theta}_{pr}}{\delta {P}_{st}}\bigg{|}_{\mathbf{P}^{\circ}} \nonumber \\
&\approx & \int \mathrm{d}^3r~\mathrm{d}^3r'\phi_{r}^*(\mathbf{r})\phi_{p}(\mathbf{r})\mathbbm{f}_{\text{Hxc}\theta}(\mathbf{r},\mathbf{r'}) \phi_{t}(\mathbf{r}')\phi^*_{s}(\mathbf{r}')\nonumber \\
&=&
\big{(}rp|\mathbbm{f}_{\text{Hxc}\theta}|ts \big{)}.
\end{eqnarray}
The working equation would be reduced to an eigenvalue equation
\begin{eqnarray}
\sum_{st}
\bigg{[}
(\varepsilon_p-\varepsilon_r)\cdot \delta_{ps,st}
&-&(f_p -f_r)
\big{(}rp\big{|}\mathbbm{f}_{\textrm{Hxc}\theta}\big{|}ts\big{)}
\bigg{]}
\cdot
\Omega^{R}_{k,st}
\nonumber \\
&=&
\omega_k \cdot
\Omega^{R}_{k,pr},
\label{right_eigenvalue_eqn}
\end{eqnarray} 
where $\Omega^{R}_{pr}=\delta P_{pr}$ and $k$ denotes the $k$-th eigenvalue. 
This can be represented in the matrix form as Casida's equation~\cite{Casida_chapter}:
\begin{equation}
\begin{pmatrix}
\mathbf{\mathcal{\hat{A}}} & \mathbf{\mathcal{\hat{B}}} \\ \mathbf{\mathcal{\hat{B}}}^* & \mathbf{\mathcal{\hat{A}}}^*
\end{pmatrix}
\begin{pmatrix}
\mathbf{X} \\ \mathbf{Y}
\end{pmatrix} = \omega_k
\begin{pmatrix}
\mathbf{\hat{I}} & \mathbf{\hat{0}} \\
\mathbf{\hat{0}} & -\mathbf{\hat{I}}
\end{pmatrix}
\begin{pmatrix}
\mathbf{X}_k \\ \mathbf{Y}_k
\end{pmatrix},
\label{Casida-like_relation}
\end{equation} 
where $X_{k,pr}=\Omega^{R}_{k,p>r}$, $Y_{k,rp}=\Omega^{R}_{k,p<r}$, denotes upward and downward transitions, respectively. 
The coupling matrices are defined as 
\begin{eqnarray}
&& A_{pr,st}=(\varepsilon_p-\varepsilon_r)\delta_{ps}\delta_{rt} + B_{pr,ts} \\
&& B_{pr,st}= -(f_p-f_r)\big{(}rp|\mathbbm{f}_{\text{Hxc}\theta}|st \big{)}.
\end{eqnarray} 
These matrices are similar in form to those derived from conventional Casida's equation which most TDDFT works are based on~\cite{TDTAO_preamble, Casida_review_2012}. However, we consider the fractional occupation number difference ($\Delta f$) pre-factor in Eq.~\ref{Casida-like_relation}, which is equivalent to the original Casida's equation in Ref.~\citenum{Casida_chapter}. It is to be noted that the occupation numbers are explicitly sourced from GS TAO. In Eq.~\ref{right_eigenvalue_eqn}, the superscript $R$ in $\Omega^R$ implies that the eigenvectors obtained are the right eigenvectors. Using the density-density response function (Appendix~\ref{density-density_response}), an eigenvalue-like equation that is complementary to that in Eq.~\ref{right_eigenvalue_eqn} can be derived. The details are included in Appendix~\ref{TDTAO_Casida}.

\subsection{\textit{Idempotency} in TDTAO-DFT}
In KS theory, an idempotent one-electron density matrix ($\mathbf{PP}=\mathbf{P}$)~\cite{TDTAO_preamble} is derived from the single-determinant ansatz of the wavefunction, 
so for any \textit{first-order} changes of the one-electron density matrix 
\begin{eqnarray} 
\mathbf{P}^{\circ} \cdot \delta \mathbf{P} + \mathbf{\delta P} \cdot \mathbf{P}^{\circ} - \delta \mathbf{P} = \mathbf{0},
\end{eqnarray}
which when represented in terms of KS orbitals, becomes
\begin{eqnarray}
(n_{p}+n_{q}-1)\cdot \delta P_{pq} = 0, 
\end{eqnarray}
where $\{n_{p}\}$ are the integer occupation numbers (either $0$ or $1$). Within this particular condition, the conventional Casida's scheme allows transitions between only occupied ($n_i=1$) and 
virtual ($n_a=0$) orbitals. On the other hand, due to fractional occupation numbers, the one-electron density matrix in TAO-DFT violates this idempotency condition for nonvanishing $\theta$. 
Therefore, a \textit{relaxed} condition in terms of TAO orbitals is proposed as 
\begin{eqnarray}
(f_{p}+f_{q}-1)\cdot \delta P_{pq} \propto \theta,
\end{eqnarray}
where the KS limit of TDTAO-DFT is recovered for $\theta \rightarrow 0$. This condition implies that transitions with $f_{p}+f_{r}$ tending to $1$ would be dominant. These transitions require one of the $p$ and $q$ orbitals to be \textit{strongly occupied}, $1/2 \leq f_r\leq 1$, with the other \textit{weakly occupied}, $0 \leq f_r < 1/2$. More details on the \textit{relaxed idempotency} condition for TDTAO-DFT can be found in Appendix~\ref{TDTAO_relaxed_idempotency} accompanying this work. 

\section{Computational details}
We implement this formalism in the development version of \textsf{Q-Chem 5.2}~\cite{shao2015advances}. All numerical results are calculated with cc-pVDZ basis set, which was determined by performing a comprehensive convergence test of different sets. The two-electron integrals are evaluated with the standard quadrature grid EML(50,194)~\cite{gill1993standard}, consisting of 50 Euler-Maclaurin~\cite{murray1993quadrature} radial grid points and 194 Lebedev~\cite{lebedev1999quadrature} angular grid points.

\section{H\textsubscript{2} bond dissociation using TDTAO-DFT}
We demonstrate how some of the challenges plaguing TDDFT are rectified with our method through the GS bond dissociation process of the H$_{2}$ molecule. This system has been studied 
extensively for many years using a plethora of methods. Successfully capturing the mechanism of bond dissociation within the framework of DFT has been elusive owing to the lack of incorporation of non-dynamical correlation effects. Within TAO-DFT, however, this challenge was resolved by choosing an appropriate $\theta$ of 40 mHartree~\cite{TAODFT_1, TAODFT_2}. It was further shown that, at the bond dissociation limit, the multi-reference character was more pronounced~\cite{TAODFT_1, TAODFT_2}. 

\begin{figure}[!h]
\centering
\includegraphics[width=0.95\columnwidth]{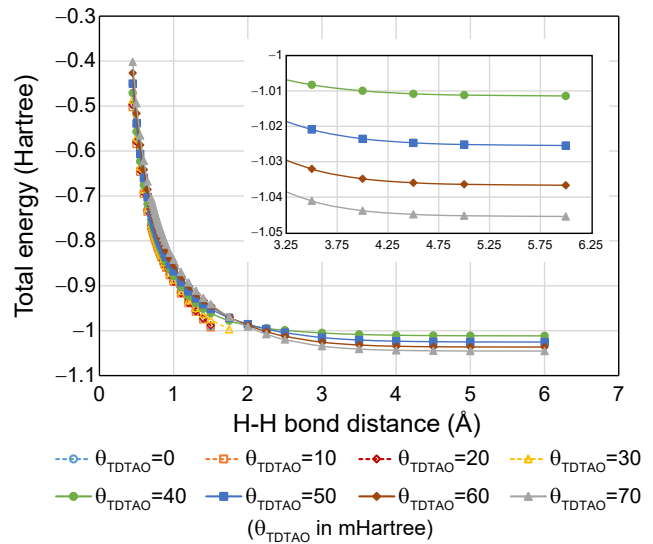}
\caption{Potential energy surface of the first triplet excited state (1\textsuperscript{3}$\Sigma_u^+$) computed using TDDFT ($\theta=0$~mHartree) and TDTAO-DFT with the PBE XC-functional, cc-pVDZ basis set and GEA version of $E_\theta$ functional~\cite{TAODFT_2} for TAO calculations. The inset shows a zoomed-in view for the large bond-distance regime.} 
\label{fig1}
\end{figure}

In TDDFT, one encounters the challenge of imaginary frequencies (i.e., excitation energies) for the triplet states which occurs in most of the results obtained from ALDA functionals 
(kernels)~\cite{Casida_2000, Reimers_2000, Baerends_2000}. This issue is related to the symmetry-breaking where the difference in spin densities (i.e., $\rho_\alpha - \rho_\beta$) is not equal to zero for a large interatomic distance. In other words, the unrestricted (asymmetric) solution obtained using KS-DFT becomes lower in total energy than the restricted (symmetric) one, as demonstrated by Casida \textit{et al.} using a two-level model~\cite{Casida_2000}. TAO-DFT significantly rectifies this issue, for a large enough $\theta$ value~\cite{TAODFT_1}. Fig.~\ref{fig1} shows the potential energy surface (PES) of the first triplet excited state (1\textsuperscript{3}$\Sigma_u^+$) for H$_{2}$ bond dissociation using TDTAO-DFT and TDDFT ($\theta=0$~mHartree).

The TDDFT results show imaginary frequencies beyond the H-H bond distance of $\sim1.5$~\AA. This is attributed to a poor ground-state reference, as mentioned previously, due to lack of incorporation of the non-dynamical correlation effects beyond this bond distance. In addition, this phenomenon is observed in TDTAO-DFT simulations for $\theta=0, 10, 20,$~and~$30$~mHartree. However, for $\theta \geq 40$ mHartree, the imaginary-frequency issue is resolved.

We also note here that the requirement for a real-value $1^3\Sigma_u^+$ excitation energy mandates a higher threshold value for $\theta$ than that obtained through a self-consistent scheme \cite{TAODFT_4}, which is around 15.5 mHartree. While a lower $\theta$ value is needed to describe the ground-state bond dissociation curves, our observation indicates that a higher $\theta$ value is needed for excitation properties and an optimal determination scheme for $\theta$ remains to be developed. One such direction is to include the excited state information in the post-SCF variational scheme similar to that outlined in Eq.~9 in Ref.~\citenum{Martinez2010}.

\begin{figure}[!h]
\centering
\includegraphics[width=0.95\columnwidth]{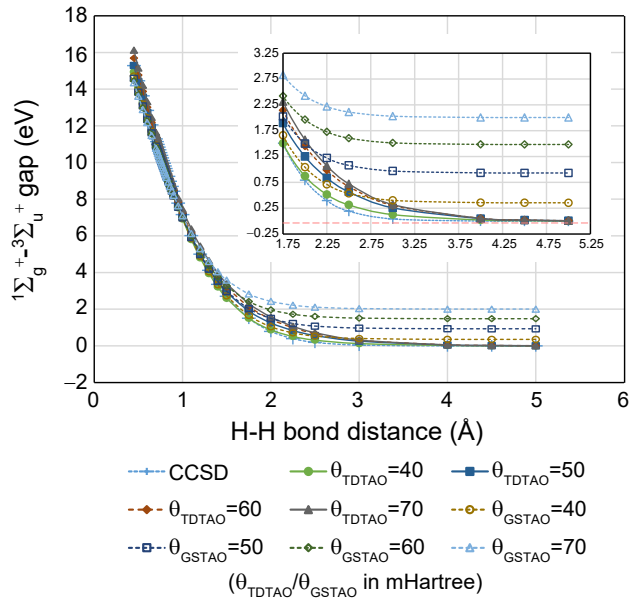}
\caption{The energy gap as a function of H-H bond distance (in \AA) between the singlet ground state (1\textsuperscript{1}$\Sigma_g^+$) and the first excited triplet state (1\textsuperscript{3}$\Sigma_u^+$) calculated using TDTAO-DFT and unrestricted TAO-DFT with the PBE XC-functional and GEA $\theta$-functional. The equation-of-motion$-$coupled cluster singles doubles (EOM-CCSD) results are presented as a benchmark. The cc-pVDZ basis set was employed for all calculations. The inset shows a zoomed-in view for the large bond-distance regime.} 
\label{fig2}
\end{figure}

Another advantageous aspect of TDTAO-DFT is that the energy of the first triplet excited (1\textsuperscript{3}$\Sigma_u^+$) state in the dissociation limit correctly approaches the 
GS singlet energy. Fig.~\ref{fig2} shows the singlet-triplet (1\textsuperscript{1}$\Sigma_g^+$-1\textsuperscript{3}$\Sigma_u^+$) vertical gap as a function of H-H bond dissociation computed using ground-state TAO-DFT, CCSD, and TDTAO-DFT. To compute the 1\textsuperscript{1}$\Sigma_g^+$-1\textsuperscript{3}$\Sigma_u^+$ gap at the ground-state level (in order to mitigate the problem of imaginary frequencies in TDDFT), it was recommended to use the unrestricted ground-state SCF formalism for H$_{2}$ and other small 
molecules~\cite{Reimers_2000, jing_kong_1, jing_kong_2}. However, this does not guarantee the convergence of the energy of the 1\textsuperscript{3}$\Sigma_u^+$ state to that of the 1\textsuperscript{1}$\Sigma_g^+$ state at the bond dissociation limit for H$_{2}$ for TAO-DFT (Fig.~\ref{fig2}). 
This gap may violate the \textit{covalent nature} of the \textsuperscript{3}$\Sigma_u^+$ state, where the energies of covalent states 1\textsuperscript{3}$\Sigma_u^+$ and 
GS (1\textsuperscript{1}$\Sigma_g^+$) should be the same at the bond dissociation limit~\cite{Baerends_2000}. In other words, at this limit, the electrons are located in the 1\textit{s} orbitals of the corresponding atoms and are therefore, isolated enough with respect to one another. This gap increases with $\theta$ due to the increase in the energy of 1\textsuperscript{3}$\Sigma_u^+$ and a simultaneous decrease in the energy of 1\textsuperscript{1}$\Sigma_g^+$ (this $\theta$-dependent decrease is also observed for the total energy of 1\textsuperscript{3}$\Sigma_u^+$ calculated with TDTAO-DFT in Fig.~\ref{fig1}). On the other hand, the trend obtained for TDTAO-DFT (Fig.~\ref{fig2}) is in excellent agreement with that obtained using the equation-of-motion coupled-cluster singles and doubles (EOM-CCSD) method or observed in experiments~\cite{kouchi1997dissociation}. EOM-CCSD is used here as a benchmark method since it is equivalent to FCI for a two-electron system like \ce{H2}.

\begin{figure*}[!h] 
\centering 
\includegraphics[width=\textwidth]{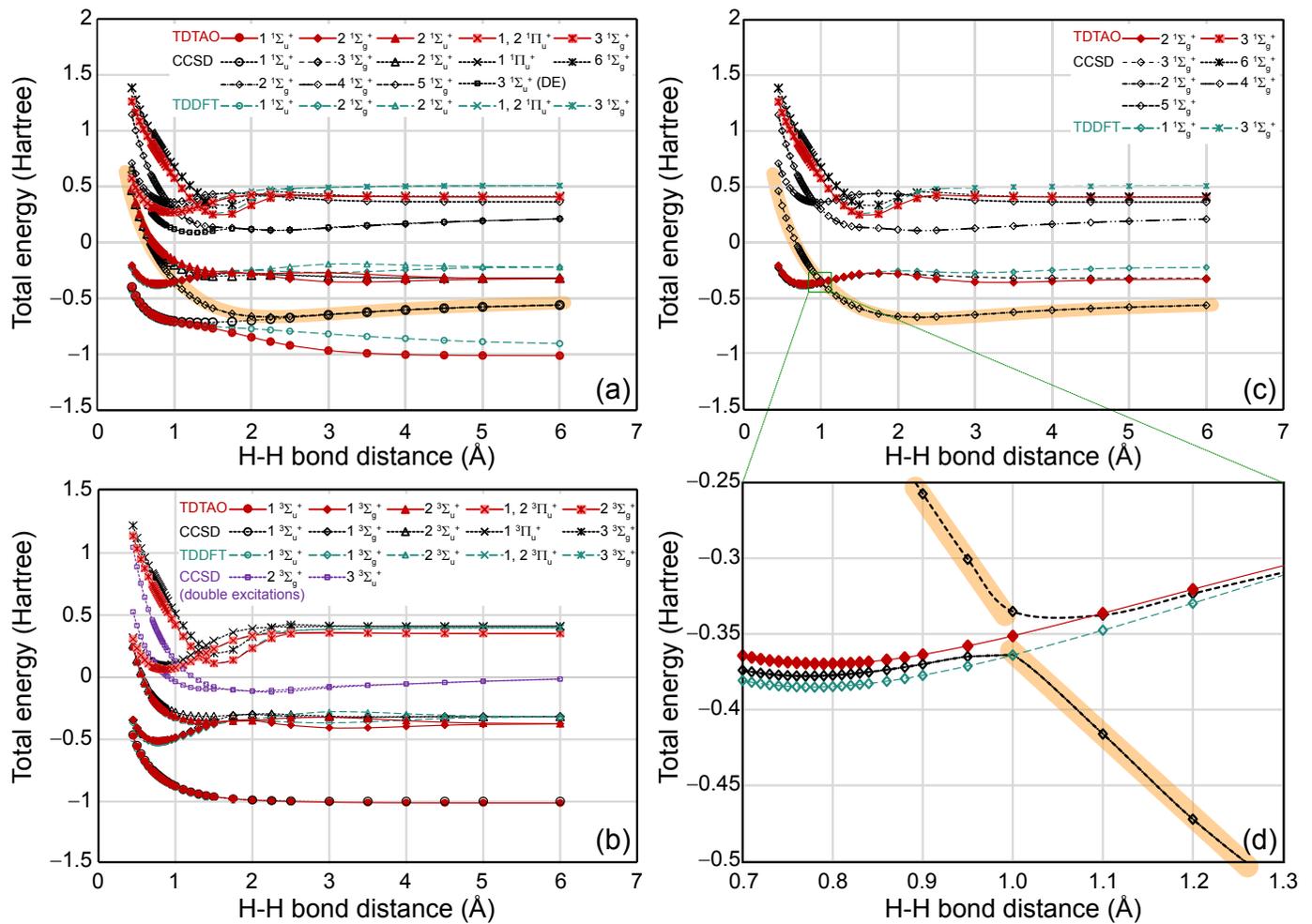} 
\caption{Potential energies of (a) singlet and (b) triplet excited states, computed using TDTAO-DFT (with $\theta$ = 40 mHartree and GEA $\theta$-functional), EOM-CCSD, and conventional TDDFT. (c) shows only the \textsuperscript{1}$\Sigma_g^+$ states. (d) is a zoomed-in region showing the avoided crossing between two EOM-CCSD states that are not completely captured by either TDTAO-DFT or conventional TDDFT. The orange shaded regions in (a), (c), and (d) indicate portions of the EOM-CCSD curves that have double excitation character. DE in (a) signifies that, that CCSD state is double excitation in nature. PBE is selected as the XC-functional for all DFT calculations and cc-pVDZ as the basis set for all calculations.} 
\label{fig3} 
\end{figure*}

For the sake of completeness, we also computed the PESs of other excited states for H$_{2}$. The lowest six singlet and triplet excited states in TDTAO-DFT and TDDFT are demonstrated with low-lying PESs from EOM-CCSD in Figs.~\ref{fig3}~(a) and (b).
The overall feature of singlet and triplet states from TDTAO-DFT is in excellent agreement with the EOM-CCSD results, except for the charge-transfer state (1\textsuperscript{1}$\Sigma_u^+$) and the missing states with double excitation character (purple curve with unfilled squares and golden yellow curves with unfilled diamonds in Figs. 3 (a) and (b)). We speculate that the problem with the 1\textsuperscript{1}$\Sigma_u^+$ state could be due to the usage of the simple adiabatic approximation to the $\text{xc}\theta$-kernel~\cite{Casida_2000, ALDA_fail_1, CT_ALDA_problem_1, CT_ALDA_problem_2} as well as the time-independent occupation numbers in our formalism~\cite{BaerendsPRL2008,BaerendsJCP2009,giesbertz2010adiabatic}. The missing CCSD double excited states also indicate the inability of TDTAO-DFT to capture the avoided crossing between the first two \textsuperscript{1}$\Sigma_g^+$ excited states (orange shaded regions as seen in Figs.~\ref{fig3} (a), (c), and (d)).
A more detailed investigation is certainly required for resolving these challenges.

\section{Relationship between $\theta$ and imaginary frequencies: A qualitative description}
\begin{figure}[!h]
\centering
\includegraphics[width=0.9\columnwidth]{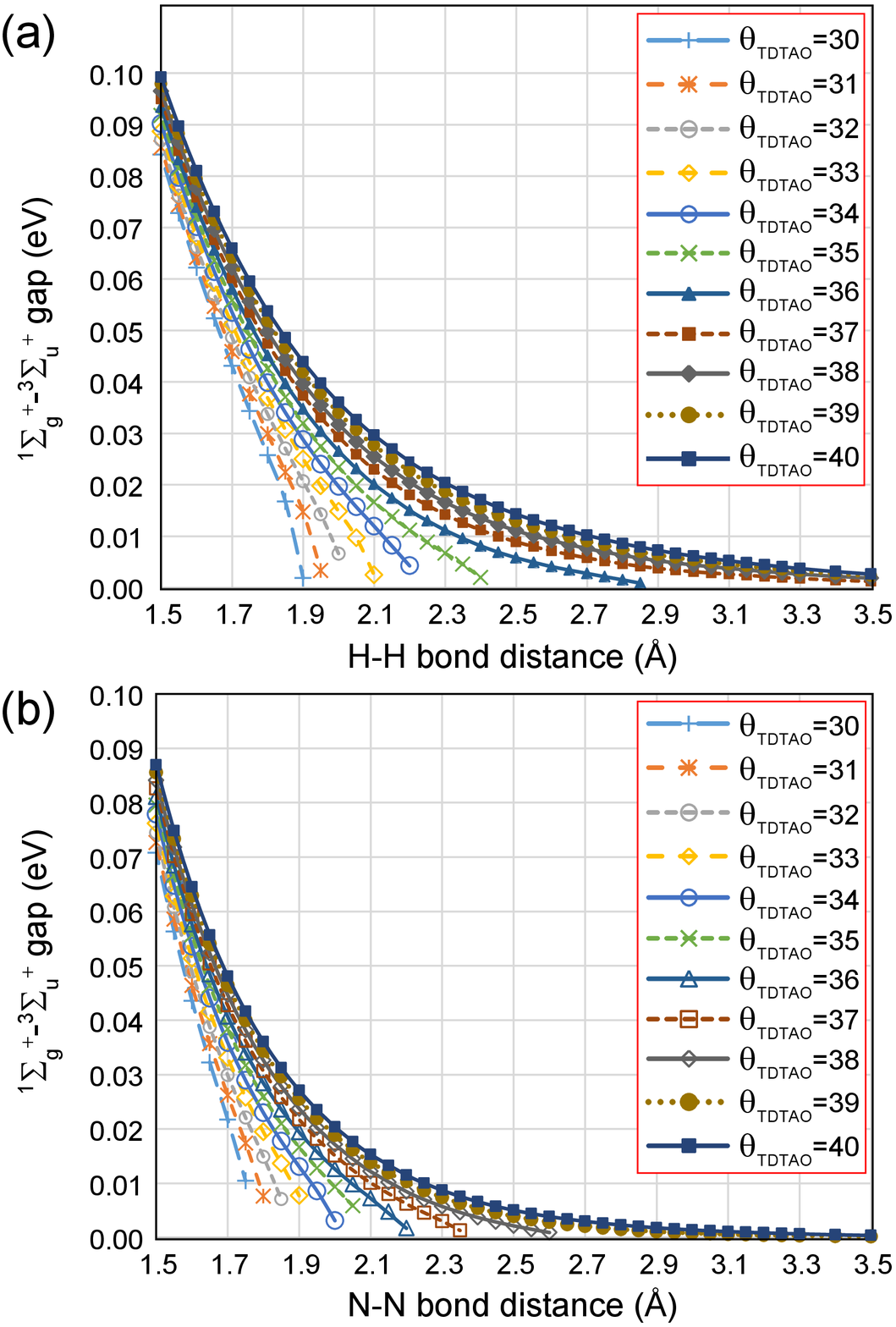}
\caption{S-T gap of (a) \ce{H2} and (b) \ce{N2} with the bond distance and different $\theta$ (in mHartree) values, calculated using TDTAO-DFT with the PBE XC-functional, cc-pVDZ basis set and GEA version of $E_\theta$ functional~\cite{TAODFT_2}. The filled symbols indicate potential energy surfaces without any imaginary frequencies.}
\label{S-T_gap_H2_N2}
\end{figure}

We perform a detailed analysis of the PESs with the different $\theta$ values to acquire more insight about the qualitative relationship between $\theta$ and the imaginary roots. Two molecular systems were chosen for this analysis, \ce{H2} and \ce{N2}, and their S-T gaps are shown in Fig.~\ref{S-T_gap_H2_N2}. The problem of imaginary frequencies is fixed with TDTAO-DFT for a suitable choice of $\theta$, irrespective of the system under consideration, thereby indicating its versatility. However, we note that $\theta$ is a system-dependent quantity and a robust algorithm is needed to ascertain it. Based on the optimal choice of $\theta$, we observe that the S-T gap vanishes at the bond dissociation limit for \ce{N2} (Fig.~\ref{S-T_gap_H2_N2}(b)), similar to that in \ce{H2} (Fig.~\ref{S-T_gap_H2_N2}(a)). This is also in agreement with experiments~\cite{Kadochnikov_2013}. 

\section{Concluding Remarks}
In summary, a time-dependent linear-response theory for predicting excited-state properties based on the TAO-DFT framework, TDTAO-DFT, is proposed. This theory takes advantage of TAO-DFT, 
where the spin-symmetry breaking problem of orbitals in ground-state SCF is resolved. As a result, TDTAO-DFT provides a correct description of low-lying triplet excited states, without imaginary energies, at the bond dissociation limit for a molecule. This was demonstrated through the dissociation curve of the hydrogen molecule, in which a reasonable lowest triplet state (1\textsuperscript{3}$\Sigma_u^+$) is captured by TDTAO-DFT, but is not so, for TDDFT. Additionally, TAO-DFT (with a large fictitious temperature $\theta$) may produce an erratic gap between the 1\textsuperscript{3}$\Sigma_u^+$ and ground states at the dissociation limit, which is resolved by TDTAO-DFT. The PESs for higher excited states of stretched \ce{H2} are also improved significantly as compared to TDDFT.

\section*{Supplementary Material}

The Supplementary Material includes additional results and the numerical data presented in this work.

\begin{acknowledgments} 

CPH acknowledges support from Academia Sinica and the \textit{Investigator Award} (AS-IA-106-M01) and the Ministry of Science and Technology of Taiwan (project 105-2113-M-001-009-MY4). 
JDC acknowledges support from the Ministry of Science and Technology of Taiwan (Grant No.\ MOST107-2628-M-002-005-MY3) and National Taiwan University (Grant No.\ NTU-CDP-105R7818). 
AM acknowledges additional financial support from the \textit{Academia Sinica Distinguished Postdoctoral Fellowship}. This work also benefited from discussions facilitated through the National 
Center for Theoretical Sciences, Taiwan. 
\end{acknowledgments} 

\section*{Data Availability Statement}

The data that supports the findings of this study are available within the article and its supplementary material.

\appendix
\section{A variational perspective of TAO-DFT}
\label{TAO-DFT_derivation}
In this the section, we briefly present the derivation of the TAO-DFT KS-like equations based on an alternative, variational principle. The same variational approach is also employed in the derivation of the linear response theory, which will be presented in the latter section of this work.  

According to the partition of energy functional~\cite{TAODFT_1,TAODFT_2}, the functional derivative of the total energy functional can be expressed as
\begin{eqnarray}
\frac{\delta E[\rho]}{\delta \phi_i(\mathbf{r})} &=&
\frac{\delta T^{\textrm{TAO}}_\mathrm{s}}{\delta \phi_i(\mathbf{r})}
+\frac{\delta V_{\textrm{ext}}}{\delta \phi_i(\mathbf{r})}
+\frac{\delta E_{\textrm{Hxc}\theta}[\rho]}{\delta \phi_i(\mathbf{r})},
\end{eqnarray}
where $T^{\textrm{TAO}}_\mathrm{s}$ is the kinetic (free) energy functional, and  $V_{\textrm{ext}}+E_{\textrm{Hxc}\theta}$ is the energy associated with the effective potential. The explicit derivative of the kinetic (free) energy functional would be 
\begin{eqnarray}
\frac{\delta T^{\textrm{TAO}}_\mathrm{s}}{\delta \phi^*_j(\mathbf{r'})}
&=&
\frac{\delta }{\delta \phi^*_j(\mathbf{r'})}\bigg{(}
\sum_i f_i  \int \mathrm{d}\mathbf{r} \,\phi^*_i(\mathbf{r}) \,\hat{t}\,\phi_i(\mathbf{r}) \nonumber \\
&&+\theta\bigg{\{ }\sum_i f_i \ln  f_i  +(1- f_i )\ln(1- f_i )\bigg{\} }
\bigg{)} \nonumber \\ 
&=&
f_j \cdot\hat{t}\,\phi_j(\mathbf{r'})
+\sum_i \bigg{[}
\frac{\delta f_i }{\delta \phi^*_j(\mathbf{r'})}\cdot
\int \mathrm{d}\mathbf{r} \,\phi^*_i(\mathbf{r}) \,\hat{t}\,\phi_i(\mathbf{r})
\bigg{]} \nonumber \\
&&+
\theta \sum_i \bigg{[}
\big{\{ }
\ln f_i-\ln(1-f_i)
\big{\} }\cdot\frac{\delta f_i }{\delta \phi^*_j(\mathbf{r'})}
\bigg{]}, 
\end{eqnarray}
where $\hat{t}=-\nabla^2/2$ and $\delta \phi_j(\mathbf{r}')/\delta \phi^*_j(\mathbf{r}') = 0$. Similarly, the derivatives of the energy term associated with external potential as well as the Hxc$\theta$ energy term are respectively,
\begin{eqnarray}
\frac{\delta V_{\textrm{ext}}}{\delta \phi^*_j(\mathbf{r'})} 
&=&
\int \mathrm{d}\mathbf{r}''
\frac{\delta \rho(r'')}{\delta \phi^*_j(\mathbf{r'})} \cdot
\frac{\delta }{\delta \rho(r'')}
\bigg{[}\int \mathrm{d}\mathbf{r}\,
v_{\textrm{ext}}(\mathbf{r})\rho(\mathbf{r})
\bigg{]}
\nonumber \\ 
&=&
\int \mathrm{d}\mathbf{r} \,
v_{\textrm{ext}}(\mathbf{r})
\frac{\delta \rho(\mathbf{r})}{\delta \phi^*_j(\mathbf{r'})}
\nonumber \\ 
&=&
f_j \cdot v_{\textrm{ext}}(\mathbf{r'}) \phi_j(\mathbf{r'}) \nonumber \\
&&+\sum_i  \bigg{(}
\frac{\delta f_i}{\delta \phi^*_j(\mathbf{r'})} \cdot
\int \mathrm{d}\mathbf{r} \,v_{\textrm{ext}}(\mathbf{r})
\phi^*_i(\mathbf{r})\phi_i(\mathbf{r})
\bigg{)},\nonumber \\
\label{vext_tao}
\end{eqnarray}
and
\begin{eqnarray}
\frac{\delta E_{\textrm{Hxc}\theta}[\rho]}{\delta \phi^*_j(\mathbf{r'})} 
&=&
\int \mathrm{d}\mathbf{r} \,
\frac{\delta \rho(\mathbf{r})}{\delta \phi^*_j(\mathbf{r'})} \cdot
\frac{\delta E_{\textrm{Hxc}\theta}[\rho]}{\delta \rho(\mathbf{r})} \nonumber \\
&=&
\int \mathrm{d}\mathbf{r} \,
\frac{\delta \rho(\mathbf{r})}{\delta \phi^*_j(\mathbf{r'})} 
\cdot v_{\textrm{Hxc}\theta}[\rho](\mathbf{r}) 
\nonumber \\
&=&
f_j \cdot v_{\textrm{Hxc}\theta}(\mathbf{r'}) \phi_j(\mathbf{r'})\nonumber \\
&&+\sum_i \Big{[}
\frac{\delta f_i}{\delta \phi^*_j(\mathbf{r'})}
\int \mathrm{d}\mathbf{r}v_{\textrm{Hxc}\theta}[\rho](\mathbf{r})
\phi^*_i(\mathbf{r})\phi_i(\mathbf{r})
\Big{]}. \nonumber \\
\label{vhxctheta}
\end{eqnarray}
Combining the three terms above, an explicit expression of the total energy functional is derived
\begin{eqnarray}
\frac{\delta E[\rho]}{\delta \phi_j^*(\mathbf{r'})} 
&=&
\frac{\delta T^{\textrm{TAO}}_\mathrm{s}}{\delta \phi_j^*(\mathbf{r'})}
+\frac{\delta V_{\textrm{ext}}}{\delta \phi_j^*(\mathbf{r'})}
+\frac{\delta E_{\textrm{Hxc}\theta}[\rho]}{\delta \phi_j^*(\mathbf{r'})} 
\nonumber \\
&=&
f_j \cdot\hat{t}\,\phi_j(\mathbf{r'})
+\sum_i \Big{[}
\frac{\delta f_i }{\delta \phi^*_j(\mathbf{r'})}\cdot
\int \mathrm{d}\mathbf{r} \,\phi^*_i(\mathbf{r}) \,\hat{t}\,\phi_i(\mathbf{r})
\Big{]} \nonumber \\
&&+
\theta \sum_i \Big{[}
\big{(}
\ln f_i-\ln(1-f_i)
\big{)}\cdot\frac{\delta f_i }{\delta \phi^*_j(\mathbf{r'})}
\Big{]} 
\nonumber \\
&&+
f_j \cdot v_{\textrm{ext}}(\mathbf{r'}) \phi_j(\mathbf{r'}) \nonumber \\
&&+\sum_i  \Big{[}
\frac{\delta f_i}{\delta \phi^*_j(\mathbf{r'})} \cdot
\int \mathrm{d}\mathbf{r} \,v_{\textrm{ext}}(\mathbf{r})
\phi^*_i(\mathbf{r})\phi_i(\mathbf{r})
\Big{]}
\nonumber \\
&&+
f_j \cdot v_{\textrm{Hxc}\theta}(\mathbf{r'}) \phi_j(\mathbf{r'}) \nonumber \\
&&+\sum_i \Big{[}
\frac{\delta f_i}{\delta \phi^*_j(\mathbf{r'})} \cdot
\int \mathrm{d}\mathbf{r}  \,v_{\textrm{Hxc}\theta}[\rho](\mathbf{r})
\phi^*_i(\mathbf{r})\phi_i(\mathbf{r})
\Big{]}
\nonumber \\
&=&
f_j \cdot \big{[}
\hat{t}+v_{\textrm{ext}}(\mathbf{r'}) + v_{\textrm{Hxc}\theta}[\rho](\mathbf{r'}) \big{]}\phi_j(\mathbf{r'}) 
\nonumber \\
&& +
\sum_i \frac{\delta f_i}{\delta \phi^*_j(\mathbf{r'})}\cdot
\bigg\{
\theta
\ln \bigg{(}\frac{f_i}{1-f_i}\bigg{)} \nonumber \\
&& +
\int \mathrm{d}\mathbf{r}\,
\phi^*_i(\mathbf{r})
\Big{[}
\hat{t}+v_{\textrm{ext}}(\mathbf{r}) + v_{\textrm{Hxc}\theta}[\rho](\mathbf{r}) \Big{]}
\phi_i(\mathbf{r})
\bigg\}. \nonumber \\
\end{eqnarray}
Enforcing the normalization conditions for both density and orbital functions, a Lagrangian is introduced
\begin{eqnarray}
\mathcal{L}[\rho] 
=
E[\rho] &-& \sum_{ij}\Big{[} \lambda_{ij}\int \mathrm{d}r \, \phi_i^*(\mathbf{r})\phi_j(\mathbf{r}) 
-\delta_{ij}\Big{]} \nonumber \\
&-&\mu\Big{[} \int \mathrm{d}\mathbf{r} \, \rho^\mathrm{TAO}(\mathbf{r}) -N_\mathrm{e}\Big{]},
\end{eqnarray}
where $\{\lambda_{ij}\}$ and $\mu$ are Lagrange multipliers. Considering the functional derivative with respect to orbital functions
\begin{eqnarray}
\frac{\delta \mathcal{L}[\rho]}{\delta \phi_j^*(\mathbf{r'})} 
&=&
f_j \cdot \big{[}
\hat{t}+v_{\textrm{ext}}(\mathbf{r'}) + v_{\textrm{Hxc}\theta}[\rho](\mathbf{r'}) \big{]}\phi_j(\mathbf{r'}) 
\nonumber \\
&&
+\sum_i  
\frac{\delta f_i}{\delta \phi^*_j(\mathbf{r'})} \cdot
\Bigg\{
\theta
\ln \bigg{(}\frac{f_i}{1-f_i}\bigg{)}
\nonumber \\
&&+
\int \mathrm{d}r\,
\phi^*_i(\mathbf{r})
\big{[}
\hat{t}+v_{\textrm{ext}}(\mathbf{r}) + v_{\textrm{Hxc}\theta}[\rho](\mathbf{r}) \big{]}
\phi_i(\mathbf{r}) 
\Bigg\}
\nonumber \\
&&
-\sum_{i}\lambda_{ji}\cdot\phi_j(\mathbf{r'})
-\mu\sum_i
\frac{\delta f_i}{\delta \phi^*_j(\mathbf{r'})}
-\mu
f_j\cdot \phi_j(\mathbf{r'})
\nonumber \\ 
&=&
f_j \cdot \big{[}
\hat{t}+v_{\textrm{ext}}(\mathbf{r'}) + v_{\textrm{Hxc}\theta}[\rho](\mathbf{r'}) \big{]}\phi_j(\mathbf{r'}) 
\nonumber \\
&&+\sum_i  \bigg\{
\varepsilon_i 
-\theta
\big{[}\frac{1}{\theta}(\varepsilon_i-\mu)\big{]}
\bigg\}
\cdot \frac{\delta f_i}{\delta \phi^*_j(\mathbf{r'})} 
\nonumber \\
&&
-\sum_{i}\lambda_{ji}\cdot\phi_i(\mathbf{r'}) 
-\mu\sum_i
\frac{\delta f_i}{\delta \phi^*_j(\mathbf{r'})}
-\mu
f_j\cdot \phi_j(\mathbf{r'})
\nonumber \\
&=&
f_j \cdot \big{[}
\hat{t}+v_{\textrm{ext}}(\mathbf{r'}) + v_{\textrm{Hxc}\theta}[\rho](\mathbf{r'}) \big{]}\phi_j(\mathbf{r'}) 
\nonumber \\
&&+
\mu\sum_i  \frac{\delta f_i}{\delta \phi^*_j(\mathbf{r'})} 
-\sum_{i}\lambda_{ji}\cdot\phi_i(\mathbf{r'}) \nonumber \\
&&-\mu\sum_i
\frac{\delta f_i}{\delta \phi^*_j(\mathbf{r'})}
-\mu
f_j\cdot \phi_j(\mathbf{r'})
\nonumber \\
&=&
f_j \cdot \big{[}
\hat{t}+v_{\textrm{ext}}(\mathbf{r'}) + v_{\textrm{Hxc}\theta}[\rho](\mathbf{r'}) \big{]}\phi_j(\mathbf{r'})
\nonumber \\ 
&&-\Big{(}\lambda_{jj}+\mu f_j\Big{)}\cdot\phi_j(\mathbf{r'})
+\sum^{i\neq j}_{i}\lambda_{ji}\cdot\phi_i(\mathbf{r'}).
\label{tao_lagrangian}
\end{eqnarray}
and using the variational condition $\delta \mathcal{L}[\rho]/\delta \phi_j^*(\mathbf{r'}) =0$, one obtains
\begin{eqnarray}
&&\big{[}
\hat{t}+v_{\textrm{ext}}(\mathbf{r'}) + v_{\textrm{Hxc}\theta}[\rho](\mathbf{r'}) \big{]}\phi_j(\mathbf{r'}) 
\nonumber \\
&& \,\,\, =  (f_j^{-1}\lambda_{jj}+\mu)\cdot\phi_j(\mathbf{r'}) + f_j^{-1}\sum_{i}^{i\neq j}\lambda_{ji}\cdot\phi_i(\mathbf{r'}).
\end{eqnarray}
Note that the second term in Eq.~\ref{tao_lagrangian} indicates that it is necessary to introduce the entropy term $\theta[\sum_i f_i \ln  f_i  +(1- f_i )\ln(1- f_i )]$ to the kinetic functional, in order to preserve the correct variational property such that the derivative terms arising from $v_\text{ext}$ and $v_{\text{Hxc}\theta}$ (last terms in Eqs.~\ref{vext_tao} and~\ref{vhxctheta}) are compensated.

With canonical orbital assumption (because the orbitals are orthonormal to one another), the equation can be recast into an eigenvalue equation, similar to a KS-like equation
\begin{eqnarray}
\hat{h}^{\textrm{TAO}}[\rho](\mathbf{r})\phi_i(\mathbf{r})
= \varepsilon_i \cdot  \phi_i(\mathbf{r}),
\end{eqnarray}
where $\hat{h}^{\textrm{TAO}}=\hat{t}+v_{\textrm{ext}} + v_{\textrm{Hxc}\theta}$ and $\varepsilon_i=f_i^{-1}\lambda_{ii}+\mu$.

\section{Detailed derivation of LR-TDTAO-DFT}
\subsection{Variational principle for TAO action functional and TD effective potential}
\label{TDTAO_variational_principle}

Starting from the action variational principle~\cite{GROSS1990255} and its modified form~\cite{Vignale}, we have the general definitions of action functionals for a physical system,
\begin{eqnarray}
&&A[\rho]
=\int^{\tau}_0 \mathrm{d}t\,
\Big\langle\Psi(t)\Big{|}
\bigg{(}
\frac{\partial}{\partial t} -\hat{H}
\bigg{)}
\Big{|}\Psi(t)\Big\rangle,
\\
&&B[\rho]
= A[\rho] + \int \mathrm{d}t \int\mathrm{d}\mathbf{r}\,
v_{\mathrm{ext}}(\mathbf{r},t)\rho(\mathbf{r},t),
\\
&&\frac{\delta B [\rho]  }{\delta \rho(\mathbf{r},t)}
=
v_{\mathrm{ext}}(\mathbf{r},t)+i\Bigg\langle\Psi[\rho](\tau)\Bigg{|}\frac{\delta\Psi_[\rho;\tau]}{\delta\rho(\mathbf{r},t)}\Bigg\rangle,
\end{eqnarray}
where $\Psi[\rho;\tau]$ represents the wavefunction at time  $t$ and $\tau$ denotes the upper bound of time-integral.

For a TDTAO system, the definition of universal action functionals can be written similarly, following that of the conventional TDDFT scheme~\cite{Vignale},
\begin{eqnarray}
&&\frac{\delta A_{\textrm{TAO}} [\rho]  }{\delta \rho(\mathbf{r},t)} = i\Bigg\langle\Psi_{\mathrm{TAO}}[\rho](\tau)\Bigg{|}\frac{\delta\Psi_{\mathrm{TAO}}[\rho;\tau]}{\delta\rho(\mathbf{r},t)}\Bigg\rangle
\\
&&B_{\textrm{TAO}}[\rho]
= A_{\textrm{TAO}}[\rho] + \int^{\tau}_0 \mathrm{d}t\mathrm{d}\mathbf{r}\,
v_{\mathrm{eff}}(\mathbf{r},t)\rho(\mathbf{r},t).
\end{eqnarray} 
The TD effective potential for TAO can be expressed as 
\begin{eqnarray}
v_{\mathrm{eff}}(\mathbf{r},t) = \frac{\delta B_{\textrm{TAO}} [\rho]  }{\delta \rho(\mathbf{r},t)}
+i\Bigg\langle\Psi_{\mathrm{TAO}}[\rho](\tau)\Bigg{|}\frac{\delta\Psi_{\mathrm{TAO}}[\rho;\tau]}{\delta\rho(\mathbf{r},t)}\Bigg\rangle.
\nonumber \\
\end{eqnarray} 
One can define the difference between the two functionals as
\begin{eqnarray}
A_{\textrm{Hxc}\theta}[\rho]=B_{\textrm{TAO}}[\rho] -B[\rho],
\end{eqnarray}
which is the TAO extension of Hartree-exchange-correlation functionals.
Summarizing the equations above, similar to TDDFT, one can recast the effective potential in TAO as 
\begin{eqnarray}
v_{\mathrm{Hxc}\theta}(\mathbf{r},t) 
&=&
\frac{\delta A_{\textrm{Hxc}\theta}[\rho] }{\delta \rho(\mathbf{r},t)}
\nonumber \\
&=&
v_{\textrm{eff}}(\mathbf{r},t) +i\Bigg\langle\Psi_{\mathrm{TAO}}[\rho](\tau)\Bigg{|}\frac{\delta\Psi_{\mathrm{TAO}}[\rho;\tau]}{\delta\rho(\mathbf{r},t)}\Bigg\rangle
\nonumber \\
&&-v_{\mathrm{ext}}(\mathbf{r},t)
-i\Bigg\langle\Psi[\rho](\tau)\Bigg{|}\frac{\delta\Psi[\rho;\tau]}{\delta\rho(\mathbf{r},t)}\Bigg\rangle
.
\end{eqnarray}

\subsection{Density-density response function}
\label{density-density_response}
Here we show that the linear response equation can also be constructed \textit{inversely}, 
\begin{eqnarray}
\delta \rho(\mathbf{r}\,t)
=
\int \textrm{d}\mathbf{r'}\textrm{d}t'\,\chi_{\textrm{s}}^{\mathrm{TAO}}(\mathbf{r}\,t,\mathbf{r}'\,t')\, 
\delta v^{\textrm{TAO}}_{\textrm{eff}}(\mathbf{r}'\,t'),
\label{LR_relation}
\end{eqnarray}
where
\begin{eqnarray}
\chi_{\textrm{s}}^{\mathrm{TAO}}(\mathbf{r}\,t,\mathbf{r}'\,t')
\equiv
\frac{\delta \rho^{\mathrm{TAO}}(\mathbf{r},t) }{\delta v^{{\mathrm{TAO}}}_{\mathrm{eff}}(\mathbf{r}',t')}
\end{eqnarray}
is the density-density response function for a non-interacting TAO system. With the density expression in terms of TD orbitals, one obtains
\begin{eqnarray}
\chi_{\textrm{s}}^{\mathrm{TAO}}(\mathbf{r}\,t,\mathbf{r}'\,t')
&=&
\frac{\delta \rho^{\mathrm{TAO}}(\mathbf{r},t) }{\delta v^{{\mathrm{TAO}}}_{\mathrm{eff}}(\mathbf{r}',t')}
=
\frac{\delta  \Big{[}
	\sum_p f_p \, \phi^*_{p}(\mathbf{r},t) \phi_{p}(\mathbf{r},t)
	\Big{]} }{\delta v^{{\mathrm{TAO}}}_{\mathrm{eff}}(\mathbf{r}',t')} \nonumber \\
&=&
\sum_p f_p 
\frac{\delta  \Big{[}
	\phi^*_{p}(\mathbf{r},t) \phi_{p}(\mathbf{r},t)
	\Big{]} }{\delta v^{{\mathrm{TAO}}}_{\mathrm{eff}}(\mathbf{r}',t')}
\nonumber \\
&=&
\sum_p f_p \bigg{[}
\frac{\delta  
	\phi^*_{p}(\mathbf{r},t) 
}{\delta v^{{\mathrm{TAO}}}_{\mathrm{eff}}(\mathbf{r}',t')}
\phi^{\circ}_{p}(\mathbf{r},t) \nonumber \\
&&\,\,\,\,\,\,\,\,\,\,\,\,\,\,\,\,\,\,\,\,\,+\,
\phi^{\circ*}_{p}(\mathbf{r},t)
\frac{\delta  
	\phi_{p}(\mathbf{r},t)
}{\delta v^{{\mathrm{TAO}}}_{\mathrm{eff}}(\mathbf{r}',t')}
\bigg{]},
\label{chain_rule}
\end{eqnarray}
where $\phi^{\circ}_{p}(\mathbf{r},t)$ and its complex conjugate represent the evolution of the TD orbitals in the absence of any TD perturbation (i.e., the TD external field). Applying the first-order perturbation theory, the TD orbital functions in a TD external field can be described by the following equation
\begin{eqnarray}
\phi_{p}(\mathbf{r},t) &=& 
\bigg{[}
1-i \int^{t}_{0} \mathrm{d}t'\sum_{rs}^{r\neq p}\phi^{\circ}_{r}(\mathbf{r})
\,e^{-i\varepsilon_r (t-t')} \nonumber \\
&&\,\,\,\,\,\,\times\Big{[}\int\mathrm{d}\mathbf{r}'  \,\phi^{\circ*}_{r}(\mathbf{r}')\delta v^{{\mathrm{TAO}}}_{\mathrm{eff}}(\mathbf{r}',t')\phi^{\circ}_{s}(\mathbf{r}') 
\Big{]}
e^{-i\varepsilon_s t'} \nonumber \\
&&\,\,\,\,\,\,\times\int\mathrm{d}\mathbf{r}\,\phi^{\circ*}_{s}(\mathbf{r})  \bigg{]}\phi^{\circ}_{p}(\mathbf{r})
\nonumber \\
&=&
\phi^{\circ}_{p}(\mathbf{r})
- i \,e^{-i\varepsilon_r t}
\int^{t}_{0} \mathrm{d}t'
\sum_{r,\,r\neq p}\phi^{\circ}_{r}(\mathbf{r}) \nonumber \\
&&\,\,\,\,\,\,\times\Big{[}\int\mathrm{d}\mathbf{r}'  \,\phi^{\circ*}_{r}(\mathbf{r}')\delta v^{{\mathrm{TAO}}}_{\mathrm{eff}}(\mathbf{r}',t')\phi^{\circ}_{p}(\mathbf{r}') 
\Big{]}e^{-i(\varepsilon_p-\varepsilon_r)t'}, \nonumber \\
\label{orbital_PT}
\end{eqnarray} 
and the corresponding orbital response functions are expressed explicitly in terms of initial orbitals (ground-state TAO orbitals) and orbital energies
\begin{eqnarray}
\frac{\delta  
	\phi_{p}(\mathbf{r},t)
}{\delta v^{{\mathrm{TAO}}}_{\mathrm{eff}}(\mathbf{r}',t')}
&=&
-i\Theta(t-t')e^{-i\varepsilon_r t} \nonumber \\
&&\,\,\,\,\,\,\times\sum_{r,\,r\neq p}\phi^{\circ}_{r}(\mathbf{r})
\phi^{\circ*}_{r}(\mathbf{r}')
\phi^{\circ}_{p}(\mathbf{r}') 
e^{-i(\varepsilon_p-\varepsilon_r)t'}
\nonumber \\
\frac{\delta  
	\phi^*_{p}(\mathbf{r},t)
}{\delta v^{{\mathrm{TAO}}}_{\mathrm{eff}}(\mathbf{r}',t')}
&=&
i\Theta(t-t')e^{i\varepsilon_r t} \nonumber \\
&&\,\,\,\,\,\,\times\sum_{r,\,r\neq p}\phi^{\circ*}_{r}(\mathbf{r})
\phi^{\circ}_{r}(\mathbf{r}')
\phi^{\circ*}_{p}(\mathbf{r}') 
e^{i(\varepsilon_p-\varepsilon_r)t'}. \nonumber \\
\label{orbital_response_function}
\end{eqnarray}  
Combining Eq.~\ref{chain_rule} and Eq.~\ref{orbital_response_function},
the time-domain non-interacting response function in TDTAO-DFT can be evaluated as follows
\begin{widetext}  
\begin{eqnarray}
\chi_{\textrm{s}}^{\mathrm{TAO}}(\mathbf{r}\,t,\mathbf{r}'\,t')
&=&
\sum_p f_p \Bigg\{
\bigg{[}
i\Theta(t-t')\,e^{i\varepsilon_r t}
\sum_{r,\,r\neq p}\phi^{\circ*}_{r}(\mathbf{r})
\phi^{\circ}_{r}(\mathbf{r}')
\phi^{\circ*}_{p}(\mathbf{r}') 
e^{i(\varepsilon_p-\varepsilon_r)t'}
\bigg{]}
\phi^{\circ}_{p}(\mathbf{r},t)
\nonumber \\
&&\,\,\,\,\,\,\,\,\,\,\,\,\,\,\,\,\,\,\,\,+\,
\phi^{\circ*}_{p}(\mathbf{r},t)
\bigg{[}
-i\Theta(t-t')\,e^{-i\varepsilon_r t}
\sum_{r,\,r\neq p}\phi^{\circ}_{r}(\mathbf{r})
\phi^{\circ*}_{r}(\mathbf{r}')
\phi^{\circ}_{p}(\mathbf{r}') 
e^{-i(\varepsilon_p-\varepsilon_r)t'}
\bigg{]}\Bigg\}
\\
&=&
i\sum_{pr}^{r\neq p} f_p\Theta(t-t') \Bigg\{
\phi^{\circ}_{p}(\mathbf{r})
\phi^{\circ*}_{r}(\mathbf{r})
\phi^{\circ}_{r}(\mathbf{r}')
\phi^{\circ*}_{p}(\mathbf{r}') 
e^{i(\varepsilon_r-\varepsilon_p)(t-t')}
-
\phi^{\circ*}_{p}(\mathbf{r})
\phi^{\circ}_{r}(\mathbf{r})
\phi^{\circ*}_{r}(\mathbf{r}')
\phi^{\circ}_{p}(\mathbf{r}') 
e^{-i(\varepsilon_r-\varepsilon_p)(t-t')}
\Bigg\} \nonumber 
\end{eqnarray}
\end{widetext}
Performing a Fourier transformation, the corresponding frequency-domain expression becomes

\begin{eqnarray}
\chi_{\textrm{s}}^{\mathrm{TAO}}(\mathbf{r},\mathbf{r}',\omega)
&=&
\sum_{pr}^{r\neq p} f_p \Bigg\{
\frac{
	\phi^{\circ}_{p}(\mathbf{r})
	\phi^{\circ*}_{r}(\mathbf{r})
	\phi^{\circ}_{r}(\mathbf{r}')
	\phi^{\circ*}_{p}(\mathbf{r}') 
}{\omega - (\varepsilon_r-\varepsilon_p) + i\eta} \nonumber \\
&&\,\,\,\,\,\,-
\frac{
	\phi^{\circ*}_{p}(\mathbf{r})
	\phi^{\circ}_{r}(\mathbf{r})
	\phi^{\circ*}_{r}(\mathbf{r}')
	\phi^{\circ}_{p}(\mathbf{r}')
}{\omega + (\varepsilon_r-\varepsilon_p) + i\eta}
\Bigg\}
\nonumber \\
&=&
\sum_{pr}^{r\neq p} (f_p -f_r) 
\frac{
	\phi^{\circ*}_{p}(\mathbf{r}') 
	\phi^{\circ}_{r}(\mathbf{r}')
	\phi^{\circ*}_{r}(\mathbf{r})
	\phi^{\circ}_{p}(\mathbf{r})
}{\omega - (\varepsilon_r-\varepsilon_p) + i\eta}, \nonumber \\
&&\text{ where}\, \eta \rightarrow 0.
\label{TAO_response_function}
\end{eqnarray}
We note that there are no \textit{self-transition} terms in both Eq.~\ref{TAO_response_function} and Eq.~\ref{orbital_PT}, since every TD orbital is considered as an orthonormalized function at any given instant of time. As a result, an explicit response function for a non-interacting reference system (TAO system) is obtained, and the resulting expression is similar to the conventional TDDFT~\cite{ALDA_success_2}.

\subsection{Alternative path to the Casida's equation}
\label{TDTAO_Casida}
Recall the partition of effective potential~\cite{ullrich2012time}
\begin{eqnarray}
\delta v_{\textrm{eff}}^{\textrm{TAO}}(\mathbf{r},\omega)
&=&
\delta v_{\textrm{ext}}(\mathbf{r},\omega)
+
\int \mathrm{d}\mathbf{r}_1 \,\delta\rho(\mathbf{r}_1,\omega)\cdot \mathbbm{f}_{\textrm{Hxc}\theta}(\mathbf{r},\mathbf{r}_1,\omega)
\nonumber \\
&=&
\delta v_{\textrm{ext}}(\mathbf{r},\omega)
+
\delta v_{\textrm{Hxc}\theta}[\rho](\mathbf{r},\omega).
\end{eqnarray}
where $\mathbbm{f}_{\textrm{Hxc}\theta}$ is the \textit{Fock matrix} defined in Eq. 17 in the main manuscript. Since an infinitesimal external field change is considered ($\delta v_{\textrm{ext}}(\mathbf{r}_1,\omega) \rightarrow 0$)~\cite{TDTAO_preamble, Casida_chapter},  Eq.~\ref{LR_relation} can be recast into
\begin{eqnarray}
\delta\rho(\mathbf{r},\omega)
&=&
\int \mathrm{d}\mathbf{r}_1 \int \mathrm{d}\mathbf{r}_2 \,\chi_{\textrm{s}}^{\textrm{TAO}}(\mathbf{r},\mathbf{r}_1,\omega) \nonumber \\
&&\times\delta\rho(\mathbf{r}_2,\omega)\cdot \mathbbm{f}_{\textrm{Hxc}\theta}(\mathbf{r}_1,\mathbf{r}_2,\omega).
\end{eqnarray}
If $\int \mathrm{d}\mathbf{r}\,
\mathbbm{f}_{\textrm{Hxc}\theta}(\mathbf{r}',\mathbf{r},\omega)$ is operated on both sides of the equation, one obtains an iterative formula
\begin{eqnarray}
\delta v_{\textrm{Hxc}\theta}(\mathbf{r},\omega)
&=&
\int \mathrm{d}\mathbf{r}_1\int \mathrm{d}\mathbf{r}_2
\mathbbm{f}_{\textrm{Hxc}\theta}(\mathbf{r},\mathbf{r}_1,\omega) \nonumber \\
&&\times\chi_{\textrm{s}}^{\textrm{TAO}}(\mathbf{r}_1,\mathbf{r}_2,\omega)
\delta v_{\textrm{Hxc}\theta}(\mathbf{r}_2,\omega)
\label{iterative_formula}
\end{eqnarray}
Recalling the explicit expression of non-interacting response function in Eq.~\ref{TAO_response_function}, Eq.~\ref{iterative_formula} can be reformulated into
\begin{eqnarray}
\delta v^{\textrm{Hxc}\theta}_{rs}(\omega)
&=&
\sum_{pq}^{q\neq p}
\int \mathrm{d}\mathbf{r}
\int \mathrm{d}\mathbf{r}_1 \,
\phi^{\circ*}_{r}(\mathbf{r}) 
\phi^{\circ}_{s}(\mathbf{r}) 
\phi^{\circ*}_{q}(\mathbf{r}_1) 
\phi^{\circ}_{p}(\mathbf{r}_1)
\nonumber \\
&& \,\,\,\,\,\, \times \,
\mathbbm{f}_{\textrm{Hxc}\theta}(\mathbf{r},\mathbf{r}_1,\omega) 
\bigg{[}\frac{
	(f_p -f_q) \cdot
	\delta v^{\textrm{Hxc}\theta}_{pq}(\omega)
}{\omega - (\varepsilon_q-\varepsilon_p) + i\eta}
\bigg{]}, \nonumber \\
\end{eqnarray}
where $\delta v^{\textrm{Hxc}\theta}_{rs}(\omega) = \int\mathrm{d}\mathbf{r}\,\phi^{\circ*}_{r}(\mathbf{r})\phi^{\circ}_{s}(\mathbf{r})\delta v_{\textrm{Hxc}\theta}(\mathbf{r},\omega)$ is the Hxc$\theta$ potential projected on the single-particle basis set. Similar to the derivation in main manuscript, the two-electron integral is defined as follows:
\begin{eqnarray}
&&\big{(}rs\big{|}\mathbbm{f}_{\textrm{Hxc}\theta}(\omega)\big{|}pq\big{)} \\
&&\equiv \int \mathrm{d}\mathbf{r}\,
\phi^{\circ*}_{r}(\mathbf{r}) 
\phi^{\circ}_{s}(\mathbf{r})
\int \mathrm{d}\mathbf{r}_1\,
\phi^{\circ*}_{q}(\mathbf{r}_1) 
\phi^{\circ}_{p}(\mathbf{r}_1)
\mathbbm{f}_{\textrm{Hxc}\theta}(\mathbf{r},\mathbf{r}_1,\omega). \nonumber
\end{eqnarray}
With a rescaling factor $[\omega - (\varepsilon_s-\varepsilon_r) + i\eta]$, an iterative equation in a finite basis set is obtained 
\begin{eqnarray}
&&\Big{[}\omega -(\varepsilon_s-\varepsilon_r)\Big{]}\cdot
\Omega^\mathrm{L}_{rs}(\omega) \nonumber \\
&&=
\sum_{pq}^{q\neq p}
\bigg{[}
\big{(}rs\big{|}\mathbbm{f}_{\textrm{Hxc}\theta}(\omega)\big{|}pq\big{)}
(f_p -f_q)
\bigg{]} \cdot
\Omega^\mathrm{L}_{pq}(\omega),
\end{eqnarray}
where 
\begin{eqnarray}
\Omega^\mathrm{L}_{rs}(\omega) \equiv
\frac{\delta v^{\textrm{Hxc}\theta}_{rs}(\omega)
}{\omega - (\varepsilon_s-\varepsilon_r) + i\eta}.
\end{eqnarray}
Within ALDA, the corresponding eigenvalue equation would be
\begin{eqnarray}
&&\sum_{pq}
\bigg{[}
(\varepsilon_q-\varepsilon_p) \delta_{qs}\delta_{pr}
-\big{(}rs\big{|}\mathbbm{f}_{\textrm{Hxc}\theta}\big{|}pq\big{)}
(f_q -f_p)
\bigg{]}
\Omega^\mathrm{L}_{k,pq} \nonumber \\
&&\,\,\,\,=
\omega_k \cdot
\Omega^\mathrm{L}_{k,rs},
\end{eqnarray}
where $k$ denotes the $k$-th eigenvalue. We note that this eigenvalue equation is not exactly the same as Eq.~\ref{right_eigenvalue_eqn} the main text. However, because of the transpose relation between the two matrices, they will generate the same eigenspectra.

\section{\textit{Relaxed idempotency} condition}
\label{TDTAO_relaxed_idempotency}
In conventional TDDFT, transitions between orbital are pre-selected by the idempotency condition~\cite{TDTAO_preamble}, which is derived from a single-determinant assumption, and can be formulated as
\begin{eqnarray}
(f_p+f_q-1)\delta P_{pq} = 0,
\end{eqnarray}
where $P_{pq}$ is a matrix element of transition density matrices. This condition leads to the result that only transitions between occupied and virtual orbitals would contribute to a physical (single) excitation. On the other hand, since the single-determinant assumption is removed from TAO-DFT, we consider an alternative invariant assumption based on the recurrence relation of the derivative of Fermi function
\begin{eqnarray}
\frac{\partial f_p }{\partial \varepsilon_p} =  -(f_p-f_p^2)/\theta,
\label{fermi_der}
\end{eqnarray}
or in matrix representation is
\begin{eqnarray}
\frac{\partial \mathbf{P}_0 }{\partial \mathbf{F}_0} =  -(\mathbf{P}_0-\mathbf{P}_0^2)/\theta,
\label{fermi_der1}
\end{eqnarray}
where $\mathbf{P}_0-\mathbf{P}_0^2$ on the left-hand-side implies a relaxed idempotency feature of TAO one-particle density matrix. In other words, instead of equating to zero,  $\mathbf{P}_0-\mathbf{P}_0^2$ is associated with another constant, $\theta\cdot{\partial \mathbf{P}_0 }/{\partial \mathbf{F}_0}$. To employ the relaxed condition in excited state TAO, we further assume that the simple partial derivative form would be preserved in the
TD extension of ${\partial \mathbf{P}_0 }/{\partial \mathbf{F}_0}$. Recall the total functional derivative of the density matrix
\begin{eqnarray}
\delta(\mathbf{P}-\mathbf{P}^2)
=   
\delta\mathbf{P}-\mathbf{P}_0\cdot\delta\mathbf{P}-\delta\mathbf{P}\cdot\mathbf{P}_0,
\end{eqnarray}
and combine it with Eq.~\ref{fermi_der}
\begin{eqnarray}
\delta P_{pq}- f_p\cdot\delta P_{pq}-\delta P_{pq}\cdot f_q
&=&
-(f_p+f_q-1)\delta P_{pq} \nonumber \\
&=&
-\theta \cdot \delta \bigg{[} \frac{\partial \mathbf{P} }{\partial \mathbf{F}} \bigg{]}_{pq},
\end{eqnarray}
where $\delta[{\partial \mathbf{P} }/{\partial \mathbf{F}}]_{pq}$ is not a explicit derivative and is assumed as an infinitesimal constant. Therefore, the relaxed condition is proposed as follows:
\begin{eqnarray}
(f_p+f_q-1)\delta P_{pq} \propto \theta.
\end{eqnarray}
Note that the \textit{original} idempotency condition would be  preserved when the KS limit is considered ($\theta\rightarrow0$). Based on the relaxed condition, an excitation should be dominated by those $p$ and $q$ terms with $(f_p+f_q)$ tending to 1. Therefore, to reduce the interference from spurious excitations~\cite{giesbertz2010adiabatic}, only transitions between \textit{strongly occupied} orbitals and \textit{strongly virtual} orbitals, where $(f_p+f_q-1)$ is minimized, are considered in the current version of TDTAO-DFT. The criteria to classify orbitals are 
\begin{eqnarray}
f_{p\in \text{occ.}}\geq \frac{1}{2} &\text{ or }& \varepsilon_{p\in \text{occ.}} \leq \mu,
\nonumber \\
f_{q\in \text{vir.}}\leq \frac{1}{2} &\text{ or }& \varepsilon_{q\in \text{vir.}} \geq \mu.
\end{eqnarray}

\bibliography{refs} 

\end{document}